\documentclass[twocolumn,showpacs,showkeys,preprintnumbers,prb]{revtex4}

\usepackage{amsmath}
\usepackage{amssymb}
\usepackage[dvips]{graphicx}           % Include figure files
\usepackage{bm}                        % bold math
\usepackage{color}

\newcommand{\beq}{\begin{equation}}
\newcommand{\eeq}{\end{equation}}
\newcommand{\beqar}{\begin{eqnarray}}
\newcommand{\eeqar}{\end{eqnarray}}
\newcommand{\bit}{\begin{itemize}}
\newcommand{\eit}{\end{itemize}}
\newcommand{\re}{\right}
\newcommand{\li}{\left}

\begin{document}

%\preprint{in preparation for some Phys. Rev. Journal}

\title{The Hydrogen Equation of State at High Densities}
\author{J.~Vorberger}
\affiliation{Centre for Fusion, Space and Astrophysics,
             Department of Physics, University of Warwick,
             Coventry CV4 7AL, United Kingdom}
\email{j.vorberger@warwick.ac.uk}

\author{D.O.~Gericke}
\affiliation{Centre for Fusion, Space and Astrophysics,
             Department of Physics, University of Warwick,
             Coventry CV4 7AL, United Kingdom}

\author{W.-D.~Kraeft}
\affiliation{Institut f\"ur Physik, Universit\"at Rostock,
             18051 Rostock, Germany}

%\date{}

\begin{abstract}
We use a two-fluid model combining the quantum Green's function
technique for the electrons and a classical HNC description for
the ions to calculate the high-density equation of state of
hydrogen. This approach allows us to describe fully ionized
plasmas of any electron degeneracy and any ionic coupling
strength which are important for the modelling of a variety of
astrophysical objects and inertial confinement fusion targets. 
We have also performed density
functional molecular dynamics simulations (DFT-MD) and show
that the data obtained agree with our approach in the high
density limit. Good agreement is also found between DFT-MD and
quantum Monte Carlo simulations. The
thermodynamic properties of dense hydrogen can thus be obtained
for the entire density range using only calculations in the
physical picture.
\end{abstract}

\pacs{52.25.Dg, 52.25.Kn, 52.27.Gr} %52.65Vv
\keywords{dense plasmas, equation of state, ab initio simulations}

\maketitle

%%%%%%%%%%%%%%%%%%%%%%%%%%%%%%%%%%%%%%%%%%%%%%%%%%%%%%%%%%%%%%%%%%%%%%%%%%%%%%%%
\section{Introduction}
The properties of a variety of astrophysical objects are essentially determined 
by the high-density equation of state (EOS) of hydrogen \cite{G_05,F_09}.
Hydrogen is fully ionized at these high densities independent of the
temperature. In this metallic phase, the electrons can be of arbitrary
degeneracy including the highly degenerate limit ($T \!=\! 0$) and the ions are
strongly coupled. The thermodynamics is thus strongly influenced by
the well-pronounced short-range order in the proton subsystem although the
major contributions stem from the electron gas.

Metallic hydrogen occurs in the interior of giant gas planets like Jupiter,
Saturn and similar extrasolar planets \cite{G_05,miljup,netjup}. The
temperatures encountered along the isentrope of giant planets are in the order
of a few electron volts. At densities comparable to solids and above, 
a $T \!=\! 0$ description of the electrons is possible for colder planets,
whereas temperature related corrections might be needed for hotter planets \cite{b_97}.
Higher temperatures and densities are found in white dwarf stars and the crust
of neutron stars \cite{DLFB_07,sj_99}. As most elements are fully ionized
under these conditions, matter behaves hydrogen-like and the EOS is again
determined by a combination of degenerate electrons and strongly coupled ions.
%\cite{note_1}.

Beyond astrophysical applications, the EOS of dense hydrogen is also required
to model inertial confinement fusion experiments as the compression path of the
fuel runs through the parameter space considered here \cite{lindl,hu}. Although at
much higher densities, the fully compressed DT-pellet has similar
properties when considering electron degeneracy and ionic coupling strength.

Most astrophysical objects exhibit an isentrope that covers several phases.
For example the internal structure of giant gas planets is determined by phase 
transitions as the molecular to metallic or atomic to metallic transition
in dense fluid hydrogen (sometimes named plasma phase transition \cite{ppt1,ppt2}).
After many investigations leaving the nature of this transition unclear 
\cite{weir,fortov,beule,scandolo,delany,tamblyn,VTMB_07,Holst,wpmd}, recent first principle 
simulations showed that a first order phase transition with surprisingly small volume 
change is indeed likely \cite{morales,lohore}.
%There are numerous experimental and theoretical results in support and in
%disagreement of this transition being first order %\cite{weir,fortov,beule,scandolo,delany,tamblyn,VTMB_07,Holst,wpmd}. Besides
%new experiments, the question might be resolved by {\em ab initio} simulations
%such as density functional molecular dynamics (DFT-MD), path integral Monte
%Carlo (PIMC), and coupled electron ion Monte Carlo (CEIMC). Such first
%principle techniques are fully capable to describe the
%physics needed to be taken into account.
%Indeed, in recent work such simulations were used to show that
%a first order transition with surprisingly small volume change is likely to
%occur \cite{morales,lohore}.

However, present {\em ab initio} simulations like DFT-MD, path integral Monte
Carlo (PIMC), or coupled electron-ion Monte Carlo (CEIMC)
cover a limited parameter space
only. For a consistent description of the EOS, one has to require that
i) different techniques agree in overlapping regions and
ii) the simulation data merge with well-founded theories in limiting
cases. The second point has been achieved only in the low density region where
PIMC simulations match density and fugacity expansions of the EOS 
perfectly \cite{greeneos}. 

In this paper, we focus on the high-density limit of the equation of state of fluid
hydrogen. As quantum Monte Carlo schemes are not available for very high
densities, we rely on DFT-MD simulations here. 
%A comparison of DFT-MD
%results with recent CEIMC data shows excellent agreement in
%an extended region in the metallic phase \cite{CEIMC}. 
However, first principle simulations like DFT-MD have so far been unable to provide
results that converge into the high density, i.e. Gell-Mann \& Brueckner, $T \!=\! 0$ limit. 
We resolve this issue by carefully performed DFT-MD simulations for higher densities
and by employing an analytic approach that extends the
$T \!=\! 0$ limiting law  to parameters with finite
temperatures. We can then demonstrate agreement in overlapping regions of
density, so that the goal of a combined EOS of hydrogen, that is solely
based on methods in the physical picture and covers the entire density range
for temperatures of a few electron volts, is reached.

The analytic EOS theory we apply for high densities is a two-fluid model based
on a perturbation with respect to the electron-ion interaction. It keeps all
contributions from correlations in the ion subsystem and is applicable for
arbitrary  degeneracy of the electrons. 
%The Green's function theory used to
%describe the electrons allows for finite temperatures but limits the
%interaction strength. 
Thus, our two-fluid model is valid for fully
ionized plasmas with, compared to thermal energies and/or ion-ion interactions,
weak electron-ion interactions. 

After an introduction of the model in the next
section, we present results and comparisons with data from first principle
quantum simulations. In particular, we show which steps are necessary to reach
agreement between our two-fluid model and the simulations and also give limits
for the applicability of both.

%%%%%%%%%%%%%%%%%%%%%%%%%%%%%%%%%%%%%%%%%%%%%%%%%%%%%%%%%%%%%%%%%%%%%%%%%%%%%%%%
\section{Analytical EOS Approach}
We consider fully ionized plasmas consisting of protons and electrons. To
characterize the interaction strength between the particles, we define the
classical coupling parameter
\beq
\Gamma = \frac{e^2}{d \, k_B T}
\qquad \mbox{with} \qquad
d = \left(\frac{4\pi}{3} n \right)^{-1/3} \,.
\eeq
In the quantum regime, the classical kinetic energy scale has to be replaced by
its quantum analog: $k_B T \to \frac{2}{3}\langle K_a\rangle$. The mean
kinetic energy $\langle K_a\rangle$ can be calculated via a Fermi integral
\cite{bluebook} which recovers both the classical as well as the fully
degenerate (Fermi energy) limits. Note that {\em classically} all coupling
parameters are identical for hydrogen while the electron-electron and the
electron-ion coupling are strongly reduced in plasmas with highly degenerate
electrons as $\langle K_e \rangle \!\gg\! k_B T$ holds here.
In the quantum limit, the electron coupling parameter $\Gamma_e$ 
becomes smaller than unity due to quantum degeneracy. Here, the Brueckner 
parameter $r_s=d/a_B$ is commonly used to describe the coupling strength.
%%%%%%%%%%%%%%%%%%%%%%%%%%%%%%%%%%%%%%%%%%%%%%%%%%%%%%%%%%%%%%%%%%%%%%%%%%%%%%%%
\subsection{Two-Fluid Model}
For the parameters considered here, the electron-proton interactions are weak
($\Gamma_{ei} \!\ll\! 1$) while the proton-proton coupling is usually strong
($\Gamma_{ii} \!\ge\! 1$). The weak interactions between electrons and protons
allows us to apply a Born-Oppenheimer approximation and treat the electrons in
linear response to the fields created by the ions. As a result, one can rewrite
the full two-component problem as a one-component system for the ions which
interact via effective potentials. More precisely, this will be called the
two-fluid model where the fully correlated electron and ion fluids interact
only weakly with each other. Of course, this procedure eliminates the ability
to describe bound states which is however unnecessary in the high density limit. 
Applying this two-fluid model, the pressure is given by 
\cite{Hansen, Ashcroft}
\beqar
\frac{\beta p}{n_i}&=&1+\frac{\beta}{V}\frac{d F(n_i)}{d n_i}
\label{yeos}\\
&-&\frac{1}{2}\beta n_i\int d{\bf r}\; \left[g_{ii}(r)-1\right]
\li(\frac{r}{3}\frac{\partial}{\partial r}-
n_i\frac{\partial}{\partial n_i} \re)v_{ii}^{\rm eff}(r)\,.\nonumber
\eeqar
with
\beq
F=F_{eg}+
  \frac{N_i}{2}\int \frac{d{\bf k}}{(2\pi)^3} \, v^2_{ei}(k)\, \chi_{ee}(k)\;.
\label{vnullint}
\eeq
A similar model was also used by, e.g., Chabrier \& Potekhin \cite{Chabrier1, Chabrier2}.
The first term in Eq.~(\ref{yeos}) represents the ideal contribution from the
classical ions. In the second term, the density derivative of the free energy
$F$ produces the contribution of the correlated electron gas via the free
energy of the isolated electron subsystem $F_{eg}$. The integral over the
electron density response function $\chi_{ee}$, to be taken in random phase approximation (RPA), is an electron-ion cross term that arises from the linear
response treatment of the electrons in the two-fluid description. The third
term in Eq.~(\ref{yeos}) accounts for ionic correlations described by the pair
distribution $g_{ii}$. The polarizability of the electron gas is here to be
taken into account via an effective ion-ion potential $v_{ii}^{\rm eff}$. This
potential must also be applied when calculating the ionic pair distribution.
The density derivative arises as the effective ion-ion potential is density
dependent via the screening function.

The internal energy can be calculated similarly via
\beqar
\label{yeos_U}
\frac{\beta U}{V}&=&\frac{3}{2} n_i+\frac{\beta}{V} U_0(n_i)\label{yueos}\\
&+&\frac{1}{2}\beta n_i^2\int d{\bf r}\; \left[g_{ii}(r)-1\right]
\left[ v_{ii}^{\rm eff}(r)-T\frac{\partial}{\partial
T}v_{ii}^{\rm eff}(r)\right].\nonumber
\eeqar
Again, we have first the ideal ion contribution and $U_0$ denotes the internal
energy of the electron subsystem. It may be approximated by the free energy
$F$~(\ref{vnullint}) for highly degenerate systems near $T \!=\! 0$;
otherwise it is given by $U_0=F-T\partial F/\partial T$. Ionic correlations are
accounted for by the integral term. The additional temperature derivative is
due to the fact that the effective ion-ion potential is also
temperature-dependent.

%%%%%%%%%%%%%%%%%%%%%%%%%%%%%%%%%%%%%%%%%%%%%%%%%%%%%%%%%%%%%%%%%%%%%%%%%%%%%%%%
\subsection{Properties of the Ion Subsystem}
The effective ion-ion interaction consistent with the model above is given by
\beqar
v_{ii}^{\rm eff}(k)&=&v_{ii}(k)+[v_{ei}(k)]^2\chi_{ee}(k)\,,\nonumber\\
&=&\frac{4\pi e^2}{k^2} \varepsilon^{-1}_e(k)\,.
\label{viieff}
\eeqar
Here, the electron part in the static dielectric function in RPA, 
$\varepsilon_e^{-1} \!=\! 1 \!+\! v_{ee}\chi_{ee}$, was
introduced. $v_{ee}=4\pi e^2/k^2$ is the Coulomb potential between electrons \cite{bluebook}. 
The bare Coulomb interaction between the protons $v_{ii}$ is thus
linearly screened by the electrons. More simple approximations for the
effective potential can be obtained for small $k$, where the Debye or Yukawa
potential, $v_{ii}^{\rm eff}(k) \!=\! 4\pi e^2 /(k^2 + \kappa_e^2)$ 
with $\kappa_e^2=(4e^2m_e/\pi\hbar^3)\int_0^{\infty} dp f_e(p)$, follows. 

The derivation of the two-fluid description (\ref{yeos}) and (\ref{yeos_U})
clearly shows that the ionic pair distribution $g_{ii}$ must be obtained 
from a one-component description of the ions where the forces are given by the
effective interaction $v_{ii}^{\rm eff}$. Possible methods to determine
$g_{ii}$ are classical Monte Carlo and molecular dynamics simulations or
techniques based on integral equations like the hypernetted chain equations
(HNC) \cite{at,hnc1,hnc2,wuensch,schwarz,wuensch2}. 

%%%%%%%%%%%%%%%%%%%%%%%%%%%%%%%%%%%%%%%%%%%%%%%%%%%%%%%%%%%%%%%%%%%%%%%%%%%%%%%%
\subsection{Contributions of the Electron Gas}
To describe the electron gas, we employ the quantum statistical method of
thermodynamic Green's functions \cite{greenbook,bluebook}. Its advantage is the
ability to describe systems with arbitrary temperatures including the correct
$T \!=\! 0$ physics, the transition to Boltzmann statistics, and the correct
high temperature (Debye-H\"uckel) law. Using this technique, a perturbation
expansion in the interaction strength can be established
\cite{greeneos,bluebook}. Including terms up to the second order, one obtains
\beqar
p_{ee}(T_e,\mu_e)&=&p_{e}^{id}(T_e,\mu_e)+p_{e}^{HF}(T_e,\mu_e)\nonumber\\
&&+p_{ee}^{MW}(T_e,\mu_e)+
p_{e}^{e^4n}(T_e,\mu_e)\,.\label{uecorr}
\eeqar
The terms are the ideal gas law, the Hartree-Fock (HF) quantum exchange term, 
the direct Montroll-Ward (MW) term, and quantum exchange contributions of the
second order ($e^4n$), respectively. As this perturbation expansion exists in
the grand canonical ensemble, the density $n_e$ is related to the chemical
potential $\mu_e$ via
\beq
n_e(T_e,\mu_e)=\frac{\partial p_{ee}}{\partial \mu_e}\label{chempot}\,.
\eeq
An inversion within {\em golden rule} is performed in order to obtain the
pressure as function of density \cite{greeneos}. This means that correlation
contributions of the free energy as function of density are taken to be equal
to the negative excess pressure as function of the chemical potential.

The ideal pressure is given by
\beq
p_{e}^{id}(T_e,\mu_e) = 
                   \frac{2k_BT_e}{\Lambda_e^3}\mbox{I}_{3/2}(\mu_e/k_BT_e) \,.
\eeq
Here, $\Lambda_e \!=\! \sqrt{2\pi \hbar^2/m_ek_BT_e}$ is the thermal deBroglie
wavelength and $I_{3/2}$  is the Fermi integral of order $3/2$ \cite{bluebook}.

First order exchange contributions are contained in the HF term
\beq
p_e^{HF}(T_e,\mu_e)=\frac{(2\sigma_e+1)e^2}{\Lambda_e^4}
\int\limits_{-\infty}^{\mu_e/k_BT_e}d \alpha\,\mbox{I}_{-1/2}^2(\alpha)\,,
\eeq
which is given by an integral over a Fermi integral of the order $-1/2$.

The Montroll-Ward term can be computed by a double integral over the dielectric
function of the electron gas $\varepsilon_e$
\beqar
\!\!\!
p_e^{MW}(T_e,\mu_e)&=&\frac{-1}{4\pi^3}\int\limits_0^{\infty} \! dp\,p^2\,
{\cal P}\!\!
\int\limits_{\pm 0}^{\infty}d\omega\,\coth\left(\frac{\omega}{2k_BT_e}\right)
\nonumber\\
&\times&\!\!\!\left[\arctan\frac{\mbox{Im}\varepsilon_e(p,\omega)}{\mbox{Re}\varepsilon_e(p,\omega)}
-\mbox{Im}\varepsilon_e(p,\omega)\right]\,.
\eeqar
It is consistent with the expansion (\ref{uecorr}) to use the RPA dielectric 
function $\varepsilon_e$.

The normal $e^4$ exchange term, accounting for exchange effects of second order,
can be written as an integral over Fermi functions,
$f_p \!=\! [\exp(\beta p^2/2m_e \!-\! \beta\mu_e) \!+\! 1]^{-1}$, and
Pauli-blocking factors, defined by $\bar{f}_p \!=\! [1 \!-\! f_p]$,
\beqar
p_{e}^{e^4n}(T_e,\mu_e)&=&m_e\!
\int\frac{d{\bf p}d{\bf q}_1d{\bf q}_2}{(2\pi)^9}
v_{ee}(p)v_{ee}({\bf p}+{\bf q}_1+{\bf q}_2)\nonumber\\
&\times&
\frac{f_{q_1}f_{q_2}\bar{f}_{{\bf q}_1+{\bf p}}\bar{f}_{{\bf q}_2+{\bf
p}}-f_{{\bf q}_1+{\bf p}}f_{{\bf q}_2+{\bf
p}}\bar{f}_{q_1}\bar{f}_{q_2}}
{
q_1^2+q_2^2-({\bf p}+{\bf q}_1)^2-({\bf p}+{\bf q}_2)^2}\,.\nonumber\\
&&
\end{eqnarray}
Here, $v_{ee}$ is the bare electron-electron Coulomb potential.

The expansion (\ref{uecorr}) accounts for direct correlations and dynamic
screening, incorporates collective oscillations (plasmons) as well as quantum 
diffraction and exchange in the electron subsystem. This expression 
is valid for weakly coupled electrons
of arbitrary degeneracy and includes in particular the low and high temperature
limiting cases of Debye-H\"uckel and Gell-Mann \& Brueckner, respectively
\cite{greeneos}.

Within the same quantum approach, protons can be incorporated and an EOS for
hydrogen can be calculated \cite{greeneos}. The advantage of an EOS of hydrogen
fully based on Green's functions is its capability to describe quantum effects
in the proton subsystem correctly. However, a hydrogen EOS based on 
expansion (\ref{uecorr}) is restricted to weak coupling in 
the proton subsystem as well, a limitation avoided within the two-fluid approach 
presented here.
%%%%%%%%%%%%%%%%%%%%%%%%%%%%%%%%%%%%%%%%%%%%%%%%%%%%%%%%%%%%%%%%%%%%%%%%%%%%%%%%
\subsection{Limiting Behavior and Applicability}
The condition of weak electron-ion coupling as used to derive
Eqs.~(\ref{yeos}) and (\ref{yueos}) is fulfilled not only for the high density
limit of highly degenerate electrons and strongly coupled ions, but also in
the high temperature and low density limits. Here, all interactions are weak
and of the same order and the EOS is given by the Debye-H\"uckel law
\cite{bluebook}
\beq
\beta p=\beta p^{id}+\beta p^{DH} \!=\! \sum_a n_a - \frac{\kappa^3}{24\pi} \,.
\label{dh}
\eeq
The first term is the ideal classical gas contribution, 
the second term is the Debye-H\"uckel correction determined by the total inverse
screening length $\kappa^2 \!=\! \sum_a 4\pi Z_a e_a n_a \beta$. 
The sums in Eq.(\ref{dh}) and in the definition of $\kappa$ run over all species $a=\{e,i\}$.

This limiting law is not fully reached by our two-fluid model. However, 
it deviates only by a tiny fraction of $p^{DH}/p^{2-fluid} \!=\! 16\sqrt{2}/23 \!=\! 0.99$ 
(similar for the internal energy and other thermodynamic functions). 
The slight disagreement can be traced back to the neglect of the
influence of the ions on the electrons. 
Note that this result can only be obtained if the contributions of the
electron gas are evaluated for finite temperatures and not just in the ground
state. 

Our quantum treatment via thermodynamic Green's functions also ensures that the
electron contribution reaches the correct high density $T=0$ limit. Moreover, the
electron contribution is the by far largest term in the thermodynamic functions for high densities. Thus, the two-fluid model also recovers the high density $T=0$ limit (with an error given by the ratio of electron to ion mass).

Consequently, the two-fluid model constitutes a valid approximation to the EOS
of fully ionized and weakly coupled hydrogen plasmas with arbitrarily
degenerate electrons. Both conditions are fulfilled for temperatures above
$T \!=\! 2.5 \!\times\! 10^5\,$K in the entire density range, where bound states do not occur and the coupling is sufficiently weak for the entire density range.

%%%%%%%%%%%%%%%%%%%%%%%%%%%%%%%%%%%%%%%%%%%%%%%%%%%%%%%%%%%%%%%%%%%%%%%%%%%%%%%%
\subsection{Results of the Two-Fluid Model}
\begin{figure}[t]
\includegraphics[width=0.48\textwidth,clip=true]{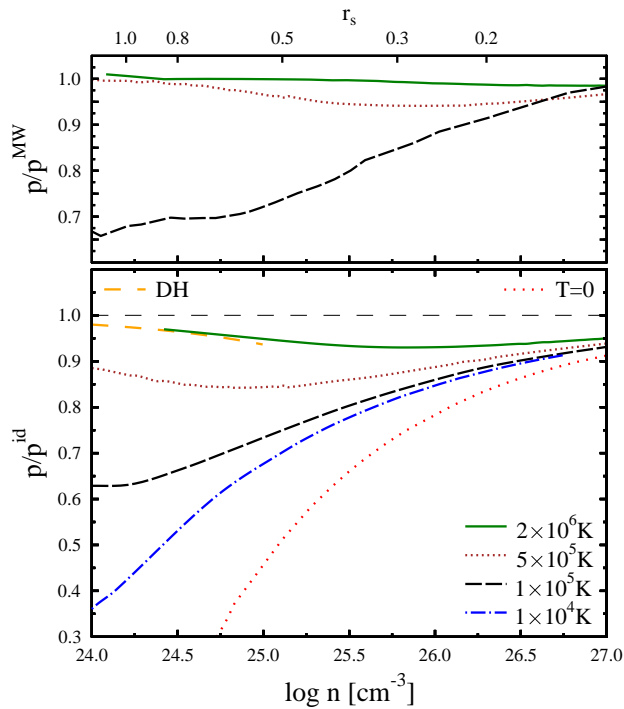}
%\plottwo{fig_01_bw.eps}{fig_01_color.eps}
\caption{(Color online) Lower panel: pressure of hydrogen as predicted by the
         two-fluid model (\ref{yeos}) normalized by the ideal
         contribution. The high temperature (Debye-H\"uckel - DH)
         and high density ($T \!=\! 0$) limits are given for
         comparison.
         Upper panel: ratio of the two-fluid model and the
         quantum statistical perturbation theory using the
         Montroll-Ward approximation for hydrogen \cite{greeneos,bluebook}.
		}
\label{pict_hd}
\end{figure}
Figure \ref{pict_hd} gives a general overview of the high-density EOS of
hydrogen as calculated within the two-fluid model. Due to the normalization by
the ideal pressure, the lines all approach unity for very high densities where
the pressure is also independent of the temperature. At intermediate densities,
the correlations yield a reduction compared to the ideal pressure. These
correlations can result from the occurrence of bound states or occur between
free particles. Bound states can be excluded for the two highest temperatures
in Fig.~\ref{pict_hd}. Still, the improved description of ion-ion correlations
within the two-fluid model yields a 10\% correction compared 
to a description of the hydrogen EOS based entirely on Green's 
function theory.

The two-fluid model of Potekhin \& Chabrier gives almost identical results to our 
approach in the parameter range where such a description can be expected to be valid, 
i.e. for metallic hydrogen and the high temperature low density case \cite{potekhin}. 
This is a nontrivial finding as the electron theories used in this paper and by 
Potekhin \& Chabrier are  quite different.

In the case of the Green's function EOS \cite{greeneos,bluebook}, the correct
high-density limit is an intrinsic feature related to the quantum treatment of
both electrons and ions. In the two-fluid model, only the electrons are treated
fully quantum statistically. Still, the two fluid system shows a similar
behavior as the high density limit is dominated by the electron contribution.
This fact holds although the ion-ion correlations strongly increases with
density since the electron terms grow significantly faster. In addition, this behavior 
also minimizes the importance of the partition function of the strongly correlated 
ion fluid. For instance, small differences between Monte Carlo and HNC treatment of 
the ions at $\Gamma_{ii}=130$ give a deviation of $2\%$ in the ion system which will 
change the result of the total equation of state by $0.16\%$.

The approach to the exact high temperature limit (here given by the
Montroll-Ward approximation of the Green's function theory) is demonstrated in
the upper panel of Fig.~\ref{pict_hd}. The two-fluid model can reproduce the
exact law within $1\%$ for $T \!=\! 2 \!\times\! 10^6\,$K. The essential 
ingredient to achieve this result is the finite temperature description of the 
electrons. Moreover, it shows that the neglected influence of the ions on the 
electrons is very weak.

%%%%%%%%%%%%%%%%%%%%%%%%%%%%%%%%%%%%%%%%%%%%%%%%%%%%%%%%%%%%%%%%%%%%%%%%%%%%%%%%
\section{DFT-MD Simulations}
DFT-MD simulations are a well-suited technique for the description of metallic
hydrogen as it is encountered for example in the inner regions of giant gas
planets. So far, there has been no overlap between an EOS calculated by
DFT-MD and {\color{red} the correct $T=0$} high-density models. To close this gap and to
allow for a direct comparison of first principle DFT-MD simulations to our
two-fluid model, we carried out DFT-MD calculations for very high densities
which require special adjustments.

We used the DFT-MD programs VASP, CPMD, and abinit
\cite{vasp1,vasp2,vasp3,cpmd,abinit1,abinit2}. The number of electrons and
protons in the simulations was $N \!=\! 128 \ldots 432$. The temperature of the
protons was controlled by a Nos\'e-Hoover thermostat \cite{nose}. The time step
in the MD simulation was chosen to be 
$\Delta t \!=\! 8\,\mbox{a.u.} \!=\! 0.194\,$fs. Every run covers at
least $2$\,ps after an initial equilibration. The exchange correlation
functional was of Perdev-Burke-Ernzerhof type \cite{pbe}.

In VASP runs, the electron-ion pseudopotential was the standard (hard)
projector augmented wave (PAW) potential as provided with the package
\cite{paw1,paw2} and used here with a plane wave cutoff of $35$\,Ha. 
For densities higher than $n \!=\! 2.2 \!\times\! 10^{24}\,$cm$^{-3}$  ($r_s \!\le\! 0.9$),
this was found to be too soft. 
Accordingly, a new harder GGA norm conserving local
pseudo\-potential was generated using the optimized pseudopotential method as included in the Opium package \cite{opium}. This new
potential was then used in the CPMD code. It has a cutoff radius of
$r_c \!=\! 0.5\,$a$_B$ ($q_c=15$, 6 Bessel functions) and requires a 
plane wave cutoff of $100\,$Ha to yield convergent results 
\footnote{A very hard pseudopotential is also used in Ref. \cite{rec},  
although with a capped energy-cutoff.}.

MD runs were usually performed using only $\Gamma$-point sampling of the
Brillouin zone. Corrections due to k-point sampling can become important in the
metallic region of hydrogen. Here, contributions of at most 2\% were determined
at randomly chosen snapshots where the pressure was reevaluated using
$2 \times 2 \times 2$ and $4 \times 4 \times 4$ Monkhorst-Pack grids of
k-points in abinit \cite{MP}. Effects due to finite electron temperatures
(Fermi smearing) have not been found for the high densities needed to achieve
overlapping with the analytical theory.

\begin{figure}[t]
\includegraphics[width=0.48\textwidth,clip=true]{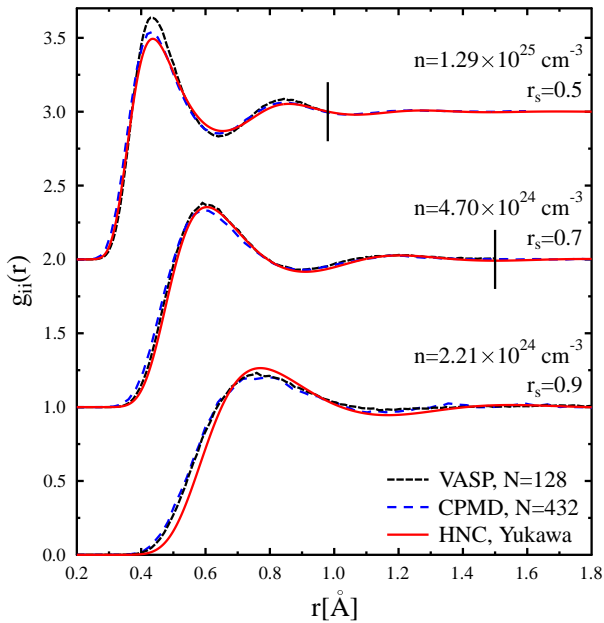}
%\plottwo{fig_02_bw.eps}{fig_02_color.eps}
\caption{(Color Online) Proton-proton pair distributions in hydrogen at
         a temperature of $T \!=\! 10^4\,$K for three
         different densities. Results of VASP (black dashed),
         CPMD (blue dashed, long spaces) and HNC using
         a linearly screened Yukawa potential (red solid) are
         compared. The black vertical lines indicate half the
         box length for the VASP runs. The lines for
         $r_s \!=\! 0.5$ and $r_s \!=\! 0.7$ have been
         shifted along the y-axis to improve readability.
}
\label{pict_gr}
\end{figure}
As a first comparison between properties of our two-fluid model and
{\em ab initio} DFT-MD simulations, Fig.~\ref{pict_gr} shows ion-ion pair
distributions. The results using VASP and CPMD deviate due to differences in
the number of particles considered and the type of pseudopotential used. In
the present study, all VASP runs have been performed with 128 electrons and
432 electrons were used in CPMD simulations. In addition, the CPMD runs use
a much harder electron-ion pseudo\-potential having a core radius of
$r_c \!=\! 0.5\,$a$_B$ whereas $r_c \!=\! 0.8\,$a$_B$ is used in the VASP runs.
For the densities with $r_s \!=\! 0.9$ and $r_s \!=\! 0.7$, these differences are 
insignificant. However, both the harder pseudopotential and the increase in the
number of particles / box size are important for the highest density with
$r_s \!=\! 0.5$. Here, VASP predicts a first correlation peak that is too high
which can be traced back to the too large core radius of the pseudopotential.
Furthermore, a box with 128 particles is too small to allow for the computation
of the structure at larger distances as it is seen in the tail of the pair
distribution.

The pair distribution functions obtained from solutions of the hypernetted
chain equations (HNC) are consistent with the approximations made to derive the
two-fluid model. The HNC solver uses a linearly screened ion-ion potential
and keeps the nonlinear correlations within the ion subsystem only. The
comparison with the DFT-MD results shows that this is an appropriate
model for the high densities shown (note that better agreement is obtained with
the CPMD data for $r_s \!=\! 0.5$). Here, the electron-ion coupling strength is
sufficiently low to justify the linear response formalism. For the lowest
density with $r_s \!=\! 0.9$, there is some deviation in the first slope and in
the strength of correlation that are a result of the linear screening
approximation used. The comparison with the DFT-MD data demonstrate that the
model of linear screening is beyond its limit of applicability for this density.

%%%%%%%%%%%%%%%%%%%%%%%%%%%%%%%%%%%%%%%%%%%%%%%%%%%%%%%%%%%%%%%%%%%%%%%%%%%%%%%%
\section{Results for the EOS}
We have already established that our two-fluid model merges with the exact
$T=0$ limiting law at the high density site of the EOS. In the following, we will
show that this model and improved DFT-MD simulations give results in agreement with 
each other over a range of densities and temperatures. As DFT-MD and further Quantum 
Monte Carlo simulations already meet the exact low density limit (Debye-H\"uckel or 
density/fugacity expansions), the entire density range can now be 
described by theories in the pysical picture.
%%%%%%%%%%%%%%%%%%%%%%%%%%%%%%%%%%%%%%%%%%%%%%%%%%%%%%%%%%%%%%%%%%%%%%%%%%%%%%%%
%\subsection{DFT-MD versus CEIMC Data}
%DFT-MD and CEIMC simulations are based on very distinct theoretical approaches
%to describe correlated quantum many-body systems. If both agree, we can state
%with high confidence that the EOS is well described by these simulation
%techniques in the tested region. This point is particularly important in
%parameter regions that cannot be described by other means. 

\begin{figure}[t]
\includegraphics[width=0.48\textwidth,clip=true]{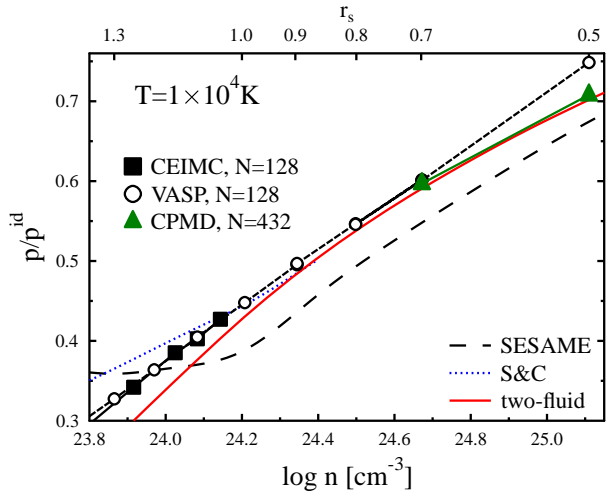}
%\plottwo{fig_03_bw.eps}{fig_03_color.eps}
\caption{(Color Online) Ratio of pressure to ideal pressure in hydrogen at
         $T \!=\! 10^4\,$K versus density. Data shown are
         obtained using VASP \cite{VTMB_07}, VASP\cite{Holst} up to $r_s=0.8$, CEIMC
         \cite{CEIMC}, Saumon \& Chabrier EOS \cite{sc1,sc2},
         and SESAME \cite{sesame}. The two-fluid model
         uses an effective ion-ion potential in RPA.
}
\label{pict_1e4}
\end{figure}
Figure \ref{pict_1e4} and Table~\ref{table1} include a comparison of EOS data
obtained by the different simulation techniques. CEIMC results were taken from
recent work \cite{CEIMC}. Unlike earlier CEIMC results \cite{CEIMC_old}, they
give almost identical values for the pressure as obtained from DFT-MD. The data
points span almost an order of magnitude in density and cover the region of
metallic hydrogen which is most difficult to describe. This excellent agreement
gives large confidence in both first principle simulations. As both data sets
have a similar slope, we can  expect that either both or none merge with the
high-density description of the two-fluid model.

%%%%%%%%%%%%%%%%%%%%%%%%%%%%%%%%%%%%%%%%%%%%%%%%%%%%%%%%%%%%%%%%%%%%%%%%%%%%%%%%
%\subsection{Two-Fluid Model versus Quantum Simulations}
\begin{table}[t]
\caption{Comparison of the hydrogen EOS as obtained by VASP
         \cite{VTMB_07,Holst}, CEIMC \cite{CEIMC}, 
         CPMD (this work, using an improved electron-proton pseudopotential) and the 
         two-fluid model (last two
         columns) using two effective proton-proton potentials
         as indicated.}
\label{table1}
\begin{tabular}{|c|c|c|c|c|c|c|}
\hline
&3500\,K&\multicolumn{5}{c|}{p/p$^{id}$}\\
$r_s$ &p$^{id}$[GPa]  & VASP & CPMD & CEIMC & ~RPA~ & Hulth\'en \\ \hline
%1.40&9.910e2&	0.236	&	-	&	0.227	&-\\
%1.30&1.430e3&	0.277	&	-	&	-	&0.215&0.250\\
%1.25&1.736e3&	-	&	-	&	0.294	&0.241&0.276\\
1.20&2.125e3&	0.321	&	-	&	-	&0.272&0.303\\
1.15&2.624e3&	-	&	-	&	0.341	&0.302&0.329\\
1.10&3.272e3&	0.369	&	-	&	0.368	&0.332&0.359\\
1.05&4.122e3&	-	&	-	&	0.393	&0.362&0.388\\
1.00&5.253e3&	0.419	&	-	&	-	&0.395&0.418\\
0.90&8.870e3&	0.472	&	-	&	-	&0.451&0.472\\
0.70&3.101e4&	-	&	0.584	&	-	&0.575&0.588\\
0.50&1.659e5&	-	&	0.702	&	-	&0.696&0.71\\\hline\hline
&$10^4\,$K&\multicolumn{5}{c|}{p/p$^{id}$}\\
$r_s$ &p$^{id}$[GPa]  & VASP & CPMD & CEIMC & ~RPA~ & Hulth\'en\\ \hline
%1.40&1.048e3&	-	&	-	&	0.283	&-	&\\
%1.30&1.497e3&	0.327	&	-	&	-	&-	&\\
%1.25&1.810e3&	-	&	-	&	0.342	&0.300&\\
1.20&2.212e3&	0.364	&	-	&	-	&0.326&0.363\\
1.15&2.721e3&	-	&	-	&	0.385	&0.348&0.384\\
1.10&3.380e3&	0.405	&	-	&	0.403	&0.377&0.404\\
1.05&4.247e3&	-	&	-	&	0.427	&0.405&0.425\\
1.00&5.400e3&	0.448	&	0.453	&	-	&0.431&0.447\\
0.90&9.060e3&	0.497	&	0.497	&	-	&0.486&0.495\\
0.80&1.623e4&	0.547	&	-	&	-	&0.538&0.547\\
0.70&3.141e4&	0.602	&	0.597	&	-	&0.589&0.602\\
0.50&1.671e5&	0.749	&	0.708	&	-
&0.700&0.713\\\hline\hline
&$2 \!\times\! 10^4\,$K&\multicolumn{5}{c|}{p/p$^{id}$}\\
$r_s$ &p$^{id}$[GPa]  & VASP & CPMD & CEIMC & ~RPA~ & Hulth\'en \\ \hline
%1.30&1.615e3&	0.383	&	-	&	-	&0.349&0.368\\
1.20&2.357e3&	0.412	&	-	&	-	&0.384&0.403\\
1.10&3.568e3&	0.445	&	-	&	-	&0.423&0.440\\
1.00&5.641e3&	0.481	&	-	&	-	&0.466&0.481\\
0.90&9.396e3&	0.522	&	-	&	-	&0.510&0.523\\
0.80&1.666e4&	0.571	&	-	&	-	&0.557&0.570\\
0.70&3.208e4&	0.619	&	0.614	&	-	&0.607&0.619\\
0.50&1.687e5&	0.761	&	0.717	&	-	&0.710&0.722\\\hline
\end{tabular}
\end{table}
Figure \ref{pict_1e4} also shows a comparison of data obtained by
{\em ab initio} simulations and results of the two-fluid model. At the low
density side where CEIMC data exist, one finds considerable deviations. These
differences are due to the decoupling of electrons and ions in the two-fluid
EOS which is not applicable here.

To test the merging of the simulation data into the two-fluid description, we
have to consider higher densities where the electron-ion coupling is weaker. For
this task, we turn to DFT-MD simulations. In the case of VASP, this is not
straightforward as one is not free in the choice of the electron-ion
pseudopotential. Particularly, for densities above $r_s \!=\! 0.7$, the particles are
closer together than the core radius of the pseudopotential. As we have already
shown for the ion-ion pair distribution (see Fig.~\ref{pict_gr}), this causes
VASP to yield unreliable results for $r_s \!\le\! 0.7$. As a result, the
pressure obtained from VASP first approaches the results of the two-fluid model
and then starts to deviate again for high densities.

The behavior of the VASP data demonstrates the need for a harder
pseudopotential with a smaller core radius. Moreover, finite size effects
become important for higher densities with larger correlation lengths. Both
issues have been resolved in CPMD runs using a new pseudopotential with a core
radius of $r_c \!=\! 0.5\,$a$_B$ and $N \!=\! 432$ electrons and protons. The
result show a smooth merging with the two-fluid model at high densities.

Figure \ref{pict_1e4} shows also results from EOS models often used in
planetary modelling namely the SESAME tables and the EOS model constructed by
Saumon \& Chabrier \cite{sc1}. The latter perfectly merges into our two-fluid model, but
deviates from the quantum simulations at lower densities. The SESAME data, on
the other hand, disagree with both {\em ab initio} simulations and the
two-fluid model over the whole density range considered. 
%These differences
%make modelling of planets containing metallic hydrogen based on SESAME highly
%questionable.
%%%%%%%%%%%%%%%%%%%%%%%%%%%%%%%%%%%%%%%%%%%%%%%%%%%%%%%%%%%%%%%%%%%%%%%%%%%%%%%%
\section{Extension and Discussion}
\subsection{Nonlinear Electron-Ion Interactions}
\begin{figure}[t]
\includegraphics[width=0.48\textwidth,clip=true]{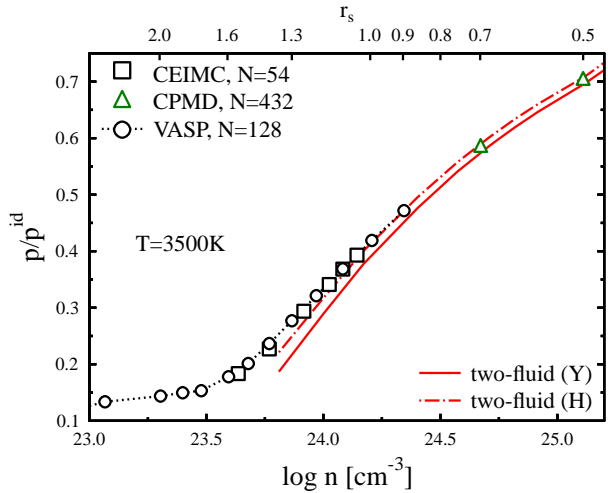}
%\plottwo{fig_04_bw.eps}{fig_04_color.eps}
\caption{(Color Online) Pressure of dense hydrogen at $T \!=\! 3500\,$K normalized
         by the ideal pressure. The symbols mark simulation data
         similar to Fig.~\ref{pict_1e4} (CEIMC data are interpolated for this temperature). 
	 The two curves show results
         of the two-fluid model with different ion-ion potentials:
         the linearly screened Yukawa potential (solid, red line)
         and the nonlinear Hulth\'en potential (red, dash-dotted line).
}
\label{pict_3500}
\end{figure}
A limitation of the two-fluid model arises from the use of first order
perturbation theory to describe the response of the electrons to the fields
created by the ions. One can however argue that the good agreement of the
two-fluid model and the fully nonlinear simulations is a result of the fact
that quadratic response terms cancel to a large extent \cite{louis}. 

An improved treatment including the fully nonlinear response (see, e.g.,
Refs.~\cite{cenni,richardson,gravel}) is beyond the scope of this paper. Here,
we estimate nonlinear effects in the electron-ion interaction by applying the
nonlinear Hulth\'en potential \cite{hulthen} 
\begin{equation}
v_{ii}^{\rm H}(r)=\frac{e^2\kappa_e}{e^{\kappa_e r}-1}
\label{viih}
\end{equation}
as an ad-hoc model for the effective ion-ion interaction.

Interestingly, the results of the two-fluid model agree rather well with the
data from the quantum simulations for $3500$K in Fig.\ref{pict_3500} if this 
nonlinear potential is used. In the density range with $1.1 \!\le\! r_s \!<\! 0.7$, the agreement is much better than for the case of the linearly screened Yukawa potential. 
Indeed, Potekhin \& Chabrier include local field corrections in the 
screening for the effective electron-proton interaction \cite{potekhin}. For the parameters of Fig.\ref{pict_3500}, our curve using the Hulth\'en potential and the curve according to their model are indistinguishable. However, there is no effect due to nonlinear electron-ion interactions for temperatures above $5000$K and densities larger than $r_s=0.9$.

%This demonstrates the need for improved analytical models in this density range if
%one does not want to rely solely on numerical simulations.

%%%%%%%%%%%%%%%%%%%%%%%%%%%%%%%%%%%%%%%%%%%%%%%%%%%%%%%%%%%%%%%%%%%%%%%%%%%%%%%%
\subsection{Quantum Effects in the Ion Subsystem}
For the highest densities considered, quantum effects may also become
important for the ions. To estimate quantum effects on the protons, we first
consider a potential that accounts for quantum diffraction effects of order
$e^2$ \cite{klimkra}
\beqar
v_{ii}^{\rm KK}(r)\!&=&\!\frac{Z^2e^2\sqrt{\pi}}{2\lambda_i\kappa_e r}
\left\{\exp\left(-\kappa_e r+\frac{\lambda_i^2\kappa_e^2}{4}
\right)\right.\label{viiq}\\
&\times&\left[\Phi\left(\frac{r}{\lambda_i}-\frac{\lambda_i\kappa_e}{2}\right)
+2\Phi\left(\frac{\lambda_i\kappa_e}{2}\right)-1\right]
\nonumber\\
\!&+&\!\left.\exp\!\left(\kappa_e r+\frac{\lambda_i^2\kappa_e^2}{4}\right)
\left[1-\Phi\left(\frac{r}{\lambda_i}+\frac{\lambda_i\kappa_e}{2}
\right)\right]\right\}.\nonumber
\eeqar
Here, $\kappa_e$ is the inverse screening length of the electrons,
$\lambda_i \!=\! \sqrt{\hbar/(2m_{ii}k_BT)}$ is the thermal wavelength of the
ions with the reduced ion mass $m_{ii} \!=\! m_i/2$, and $\Phi(x)$ denotes the
error function \cite{math}. For large distances, the 
quantum potential (\ref{viiq}) approaches the screened effective ion-ion
potential (\ref{viieff}). At the origin, it has the finite value 
\beq
\lim_{r\to 0} v_{ii}^{\rm KK}=\frac{Z^2e^2\sqrt{\pi}}{\lambda_i}
\exp\left(\frac{\lambda_i^2\kappa_e^2}{2}\right)
\left[1-\Phi\left(\frac{\lambda_i\kappa_e}{2}\right)\right] \,,
\eeq
which reflects quantum diffraction effects.

Moreover, quantum exchange effects can be important for high densities.
These can be included by adding the following exchange potential \cite{CPP_PNP}
\beqar
\label{viix}
v_{ii}^{\rm ex}(r)&=&\frac{1}{2} \exp\left(-r^2/\lambda_i^2\right) \\
&\times& \left[k_BT-\frac{Z^2e^2\pi}{4r}
\int\limits_0^1\frac{d\alpha}{\alpha} \,
\Phi\left(\frac{r\alpha}{\lambda_i\sqrt{1-\alpha}}\right)
\right] \,. \nonumber
\eeqar
The first term accounts for ideal exchange in an averaged way whereas the second
term gives the $e^2$ contribution.

The quantum potentials (\ref{viiq}) and (\ref{viix}) are derived from Slater sums, and are exact in the sense of perturbation theory. They give the correct quantum
thermodynamic functions in the weakly coupled and weakly degenerate limit,
i.e., for systems with $\Gamma_{ii} \!\ll\! 1$ and
$n\Lambda_i^3 \!=\! n(2\pi\hbar/m_ik_BT)^{3/2} \!\ll\! 1$. We use these
potentials here to estimate for which densities quantum effects start to
influence the ion properties. It should be
emphasized that the quantum potentials (\ref{viiq}) and (\ref{viix}) are used
{\em only} in the classical method employed to determine the pair distribution
function $g_{ii}$ (here, in the HNC solver). The thermodynamic
expressions (\ref{yeos}) and (\ref{yueos}) are valid in quantum as well as in
the classical case. Hence, the effective ion-ion potential (\ref{viieff}) must
be used without any quantum corrections in these formulas.

\begin{figure}[t]
\includegraphics[width=0.48\textwidth,clip=true]{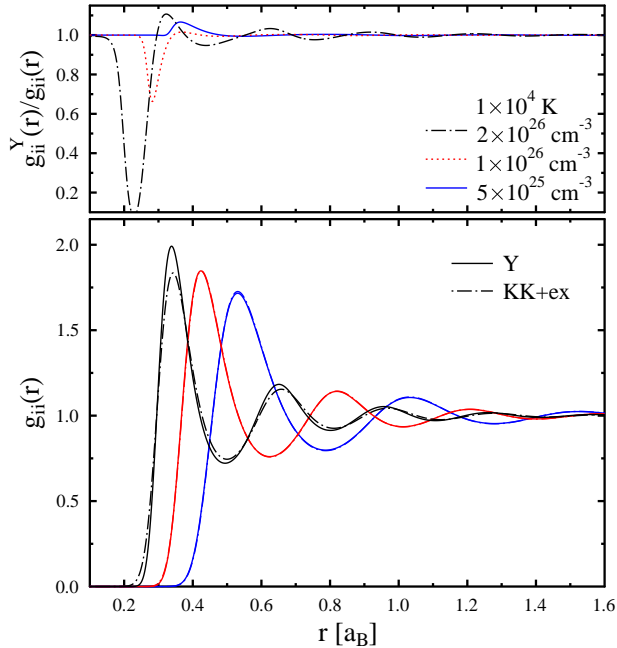}
%\plottwo{fig_05_bw.eps}{fig_05_color.eps}
\caption{(Color Online) Ion-ion pair distributions for $T \!=\! 10^4\,$K and
         three densities calculated by HNC using the screened
         potential (\ref{viieff}) (solid lines, labelled Y)
         and the screened quantum potential with exchange
         (\ref{viiq}) plus (\ref{viix}) (dash-dotted lines,
         KK+ex). The ion degeneracy is $n\Lambda_i^3 \!=\! 0.26$,
         0.52, 1.05 for the three densities, respectively. The
         ion-ion coupling is $\Gamma_{ii} \!=\! 99(96)$,
         125(119), 157(144), respectively, where the quantum
         coupling strength is given in brackets. The upper panel
         gives the ratio of the pair distributions with and
         without quantum effects.
}
\label{pict_grkkx}
\end{figure}
Figure \ref{pict_grkkx} shows ion pair distributions obtained using the usual
screened potential  (\ref{viieff}) and the quantum pseudo\-potential.
Deviations become obvious for plasmas with $n\Lambda_i^3 \!>\! 0.6$. It can be
observed that the quantum effects weaken the short range order in the proton
system. It is however remarkable that these effects only set in at these high
values of degeneracy. The reason for this behavior is found in the strong
Coulomb forces at high densities that create a large correlation hole.
Consequently, the protons are too far apart to experience short range
diffraction and quantum exchange effects. 

The emerging quantum effects in the ion structure yield only small changes in
the ion contribution to the thermodynamic functions. We determine a reduction
of less than 1\% in the correlation contribution to the ion pressure for
plasmas with $n\Lambda_i^3 \!\le\! 1$. At higher densities, the deviations from
the classical result can be large, but their exact calculation requires
PIMC methods.

The only quantum effect outside our control are zero point oscillations of the
ions. In strongly coupled liquids, caging of the ions occurs and,
for the time the cage is stable, the ions perform oscillations as in a solid \cite{donko}.
PIMC calculations in the solid and fluid phases of a Yukawa system did indeed
find contributions due to the zero point motion, but for very high densities of
$n \!=\! 10^{27}\,$cm$^{-3}$ \cite{graham}. Moreover, the overall change in
energy due to quantum oscillations is below 1\% for the parameters considered.
Furthermore, DFT-MD simulations, that do not include quantum oscillations of
the ions, and CEIMC results, that include them, agree well in the considered
parameter range which again indicates that these effects are small.

%%%%%%%%%%%%%%%%%%%%%%%%%%%%%%%%%%%%%%%%%%%%%%%%%%%%%%%%%%%%%%%%%%%%%%%%%%%%%%%%
\subsection{Hydrogen EOS for Arbitrary Densities}
The two-fluid model, based on the quantum statistical Green's function approach
for the electron contributions and classical HNC methods for the ion properties,
is well applicable in the density range that is neither covered by 
present first principle simulations nor by the $T=0$ limit. After
examining the high-density part of the EOS in detail, it is worth noting that
theories and simulations in the physical picture can be used to cover the entire 
parameter range {\em and} show agreement in overlapping regions of their applicability.

\begin{figure}[t]
\includegraphics[width=0.48\textwidth,clip=true]{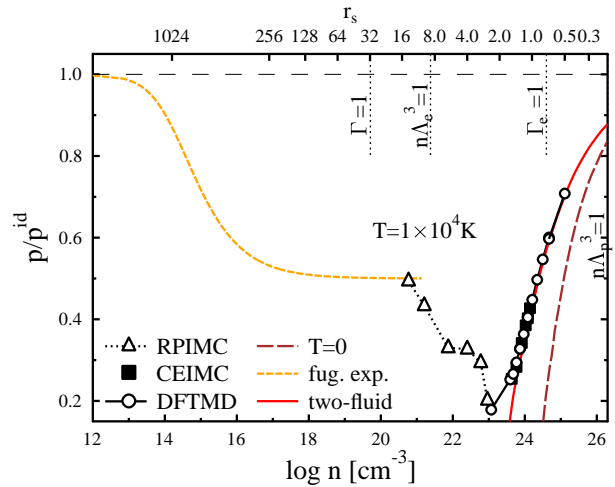}
%\plottwo{fig_06_bw.eps}{fig_06_color.eps}
\caption{(Color Online) Pressure over ideal pressure of a hydrogen plasma with
         $T \!=\! 10^4\,$K for a wide density range covering the ideal 
         classical plasma state, the atomic gas, the molecular gas, and the metallic 
         fluid (from left to right). The lines
         show the fugacity expansion from Ref.~\cite{bluebook},
         the two-fluid model (this work) using an effective
         ion-ion potential in RPA, and the $T \!=\! 0$ limit. The
         symbols mark simulation data from RPIMC \cite{PIMC},
         CEIMC \cite{CEIMC}, and DFT-MD from this work and Refs. \cite{VTMB_07,Holst}.
}
\label{pict_1e4all}
\end{figure}
Figure \ref{pict_1e4all} demonstrates this fact for an isotherm that covers the 
low density ideal plasma, the atomic fluid, the molecular fluid, and the
metallic fluid to the border of the fully degenerate electron-proton system.
Although each phase requires well-suited theories or simulations,
one finds a smooth EOS that is often based on several methods and that does not
involve any interpolation. Such a combined EOS can serve as an excellent basis
for the modelling of gas planets and other compact objects dominated by hydrogen.

%%%%%%%%%%%%%%%%%%%%%%%%%%%%%%%%%%%%%%%%%%%%%%%%%%%%%%%%%%%%%%%%%%%%%%%%%%%%%%%%
\section{Conclusions}
We have applied a two-fluid model to investigate the hydrogen EOS at high
densities. Our approach combines the advantages of a quantum theory
based on thermodynamic Green's functions for the electrons with a classical
description of the structure in the ion fluid. Thus, it is able to account for
electron degeneracy, finite temperatures, and strong forces between the ions.
Its only limitation is the requirement of weak electron-ion interactions which
naturally excludes bound states. 

The two-fluid model agrees very well with a number of exact limiting laws:
i)   at low densities it almost exactly merges with the Debye-H\"uckel law;
ii)  at high energies it practically coincides with the $T \!=\! 0$ limit as
     long as quantum effects on the proton subsystem are negligible; and 
iii) for high plasma temperatures it agrees with a perturbation expansion in
     terms of interaction strength for the entire density range.

In the high density region, we find excellent agreement of the two fluid model
with first principle simulations (DFT-MD). Our DFT-MD simulations agree also
well with recent CEIMC results. At lower densities ($r_s \!>\! 0.7$), the
request of weak electron-ion coupling is not fulfilled and one finds
increasing deviations between the results of the two-fluid model and the
quantum simulations. For densities with $r_s \!\leq\! 0.7$, the two-fluid
model is however a reliable and computationally cheap alternative to full
scale quantum simulations. From the agreement in pressure and ion structure between simulations and two-fluid model, we can conclude that the exchange correlation functional used in DFT-MD and our Green's function approach give the same electronic contributions for the relevant parameters.
The two-fluid model also bridges the parameter space where neither DFT-MD
simulations nor the $T \!=\! 0$ law can be applied. This success means that
hydrogen can confidently be described in large parts of the phase diagram
by techniques using the physical picture. For temperatures of a few electron
volts relevant for astrophysics and inertial fusion, this includes the entire density range.

Although the two-fluid model was applied here only for hydrogen, it is also
applicable for any fully ionized system with higher ion charge states or very
stable inner shell configurations with small modifications. The essential test
is if the electron-ion interaction can be considered to be small. Accordingly,
systems like the fluid region in white dwarfs can be confidently described by
the two-fluid model presented here as well.

%%%%%%%%%%%%%%%%%%%%%%%%%%%%%%%%%%%%%%%%%%%%%%%%%%%%%%%%%%%%%%%%%%%%%%%%%%%%%%%%
\section*{Acknowledgement}
The authors are grateful to B.~Holst, W.~Lorenzen, and R.~Redmer for valuable
remarks and comparisons with their DFT-MD results. We also thank M.~Schlanges
for stimulating discussions. Financial support from the UK's Engineering and
Physical Sciences Research Council is gratefully acknowledged.

%%%%%%%%%%%%%%%%%%%%%%%%%%%%%%%%%%%%%%%%%%%%%%%%%%%%%%%%%%%%%%%%%%%%%%%%%%%%%%%%

%\bibitem{Collins}
%\bibitem{dft} W. Kohn,  L. J. Sham, 
%        Phys. Rev. {\bf 140}, A1133 (1965).
%\bibitem{mermin} N. D. Mermin, 
%        Phys. Rev. {\bf 137}, A1441 (1965).

%%%%%%%%%%%%%%%%%%%%%%%%%%%%%%%%%%%%%%%%%%%%%%%%%%%%%%%%%%%%%%%%%%%%%%%%%%%%%%%%%
\end{document}